# Transverse spin-orbit interaction of light

Tong Fu[1], Jiaxin Lin[1], Yuhao Xu[1], Junji Jia[1], Yonglong Wang[2], Shunping Zhang[1, 3*], Hongxing Xu[1, 3, 4, 5 *]

[1] *Key Laboratory of Artificial Micro- and Nano-structures of Ministry of Education, and School of Physics and Technology, Wuhan University, Wuhan 430072, China.*

[2] *School of Physics and Electronic Engineering, Linyi University, Linyi 276005, China*

[3] *Wuhan Institute of Quantum Technology, Wuhan 430206, China.*

[4] *School of Microelectronics, Wuhan University, Wuhan 430072, China.*

[5] *Institute of Quantum Materials and Physics, Henan Academy of Sciences, Zhengzhou 450046, China.*

## Abstract

Light carries both longitudinal and transverse spin angular momentum. The spin can couple with its orbital counterpart via the Berry phase, known as the spin-orbit interaction (SOI) of light. The SOI of light discovered previously belongs to the longitudinal one, which relies on the Berry phase in momentum space, such as the optical Magnus effect and the spin Hall effect. Here, we show that transverse SOI, relying on the Berry phase in real space, is inherent in the Helmholtz equation when transverse spinning light propagates in curved paths. The transverse SOI lifts the degeneracy of dispersion relations of light for opposite transverse spin states, analogous to the Dresselhaus effect. Transverse SOI is ubiquitous in nanophotonic systems where transverse spin and optical path bending are inevitable. It can also explain anomalous effects like the dispersion relation of surface plasmon polariton on curved paths and the energy level of whispering gallery modes. Our results reveal the analogies of spin photonics and spintronics and offer a new degree of freedom for integrated photonics, spin photonics, and astrophysics.

## Introduction

The spin-orbit interaction (SOI) is a fundamental phenomenon that plays an important role in various subfields of physics, including atomic physics, condensed matter physics, and fluid mechanics. In electronic systems, the SOI is the coupling between the spin angular momentum (AM) and the effective magnetic field ($B_{eff}$) in both real and momentum space produced by orbital AM. For $B_{eff}$ in real space, the SOI on the one hand gives rise to the fine structure of atoms, on the other hand splits the dispersion relation of the degenerated spin states in crystals, such as the Dresselhaus effect and the Rashba effect[1,2]. For $B_{eff}$ in momentum space, the SOI produces a spin-dependent transport of electrons, known as the spin Hall effect[3]. The effective magnetic field (or Berry curvature) in momentum space is defined via the Berry phase $\Phi_G = \oiint B_{eff} \cdot da$, where the surface integral ($da$) is carried out in momentum space.

In photonic systems, the SOI of light is also general during the focusing, scattering, reflection, and refraction[4]. However, unlike electrons, electromagnetic waves can carry both longitudinal and transverse spin AM due to their vectorial nature[5-8]. The previously discovered SOI of light is mainly limited to the longitudinal SOI with a $B_{eff}$ in the momentum space, as shown in the left penal of Fig. 1. For example, the $B_{eff}$ resulted from the spin-redirection Bortolotti-Rytov-Vladimirskii-Berry phase in the momentum space can lead to the spin Hall effect[9-11] or the spin-dependent vortex generation[12-19] (Fig. 1). Other examples of longitudinal SOI, stemming from $B_{eff}$ in real space, are realized through complex design in spin photonics[20-22].

So far, transverse SOI has only been realized via the deliberate design of metamaterials[23]. Here, we show that transverse SOI is inherent in the wave equation when a transverse spinning light propagates in a curved path. Different from the longitudinal SOI, the geometric phase of transverse SOI resides in the real space where curvature acts like a gauge field and transverse spin plays the role of 'charge'. Due to the particular vector diagram between spin and orbital AM (right panel in Fig. 1), the lifting of the degeneracy of dispersion relation of the opposite transverse spin state appeared, which is analogous to the Dresselhaus effect in electronic systems. In an optical cavity, the transverse SOI shifts the energy level of whispering gallery mode (WGM), which is analogous to the fine structures of an atom.

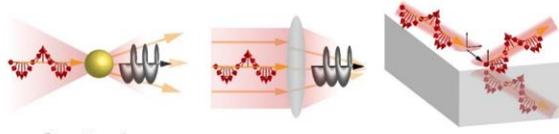

**Fig. 1. Longitudinal and transverse SOI of light.** For longitudinal SOI, the geometric phase and the effective magnetic field are in the momentum space and affect the spatial distribution. For transverse SOI, the geometric phase and the effective magnetic field are in the real space and affect the spatial distribution, momentum, and energy of the cavity photon. The yellow arrow represents the wave vector and the black and red arrows represent the orbital and spin AM, respectively.

## Results

In a homogeneous nonconducting media, Maxwell's curl equations can be combined to get the Helmholtz wave equation $(\nabla^2 + k^2)\psi = 0$, where $k = \sqrt{\varepsilon\mu}\omega$ is the wave number, and $\psi$ is the wave function of the electromagnetic field. This Helmholtz equation describes propagating electromagnetic waves in a flat space and the polarization states are tangent to the momentum sphere. Considering an electromagnetic wave propagating in a curved path with a longitudinal component of fields, the Helmholtz equation becomes[24] (More details in Supplementary Section 1)

$$\left[(\partial_t - i\kappa\hat{\sigma})^2 + \frac{\kappa^2}{4} + k^2\right]\psi = 0, \quad (1)$$

where the operator $\partial_t$ is the differential operator in a curvilinear coordinate and the subscript $t$ represents the tangent curvilinear coordinate variables. $\kappa$ is the curvature of the propagation path and $\hat{\sigma} = \begin{pmatrix} 1 & 0 \\ 0 & -1 \end{pmatrix}$ is the Pauli operator. The scalar geometrical field $\kappa^2/4$ results from the action of normal derivatives on the rescale factor. $\psi = (\psi^+, \psi^-)^T$ is the eigenstate of the transverse spin of the electromagnetic wave and T denotes the transposition. The electric field of the transverse spin eigenstate can be written as $\psi^\pm = E_t \pm iE_l$, where $E_t$ and $E_l$ are the transverse and the longitudinal components of the electric field, respectively. The most prominent change is the transverse spin-orbit coupling term $i\kappa\hat{\sigma}$ in comparison with the original Helmholtz equation in Eq. (1), i.e., $\hat{P} = \partial_t \to \partial_t - i\kappa\hat{\sigma}$. As a result, a curvature-induced geometrical phase can be obtained[24]:

$$\Phi_G = \sigma \int \kappa \, dr. \quad (2)$$

Here, $\sigma = \pm 1$ are the eigenvalues of transverse spin AM eigenstates $\psi^\pm$, and the integration is along the curvature path in real space. The geometric phase in Eq. 2 is not dependent on the direction of the path manifest the SOI preserves time-reversal symmetry. As shown in Fig. 2b, the curvature-induced geometric phase originated from the rotations of local coordinate frames with respect to the global laboratory frame when the light possesses a transverse spin[25]. The geometric phase is absent when ignoring the transverse spin[26,27]. In Fig. 2b, the red and blue arrows represent the instantaneous field in a curved and planar path, both of which possess a transverse spin. Assuming that the field in the curved path is in phase with that in the planar path initially, and the accumulated geometrical phase is exactly the rotation angle experienced when it propagates, i.e., $\Phi_G = \varphi$. The geometric phase has been numerically demonstrated in Fig. S2.

Equations 1 and 2 resemble the Schrödinger equation when describing the dynamics of a charged particle in the electromagnetic field and the corresponding geometric phase, i.e., $\hat{p} = -i\hbar\nabla \rightarrow -i\hbar\nabla - q\mathbf{A}$. The real space geometric phase of charged particles is $\Phi_G = \frac{q}{\hbar}\int \mathbf{A} \cdot d\mathbf{r}$, where $q$ is the charge of quantum particles and $\mathbf{A}$ is the gauge field[28,29]. One can find in Eq. 2, the curvature acts like a real space gauge field and transverse spin plays the role of 'charge' (Comparison shown in Fig. S3). Based on the geometric phase in Eq. 2, there is an effective magnetic field $\frac{1}{2\pi}\oiint \mathbf{B}_{\text{eff}} \cdot d\mathbf{a} = \frac{1}{2\pi}\sigma \oint \kappa dr = 1$ in a two-dimensional ring resonator in real space (Fig. 2c). The sign of the geometric phase is critically dependent on the sign of $\mathbf{s}_t \cdot \mathbf{l}_t$ (Fig. 2b), where $\mathbf{s}_t$ and $\mathbf{l}_t$ are the unit direction of transverse spin and orbital AM, respectively. In Fig. 2 and 3, light propagates in the *xy* plane so the transverse spin and orbital AM are both in the *z* direction $\mathbf{s}_t = \mathbf{s}_z$ and $\mathbf{l}_t = \mathbf{l}_z$. The transverse SOI term can be rewritten as $\kappa\hat{\sigma} = \text{sgn}(\mathbf{s}_t \cdot \mathbf{l}_t)\kappa$, where sgn is the signum function. The essence of the similarity between the transverse SOI of light and the Dresselhaus effect resides in the real space effective magnetic field[1,2].

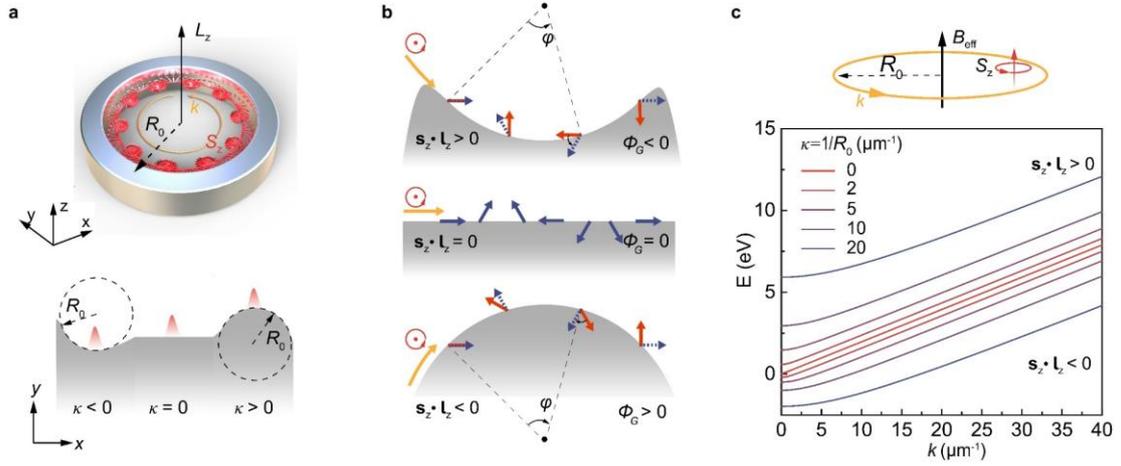

**Fig. 2. Transverse spin-orbit interaction.** (**a**) Schematic of transverse spin-orbit interaction of light and the definition of two-dimensional curvatures. (**b**) The geometric phase of transverse spin in a curved path. The yellow arrow represents the wave propagation direction; The red and blue arrows represent the instantaneous field in a curved and planar path, respectively. And the angle between the red arrows and the blue arrows denotes the geometric phase. (**c**) The dispersion relation of transverse spinning light in a curved path.

The transverse SOI is a mutual conversion between spin and orbital AM, as a result, the local transverse spin density and the dispersion relation of light in a curved path will change. Considering a transversely spinning light propagates in a rotationally symmetric curved path, as shown in Fig. 2a and 2c. The momentum is in the *xy* plane and transverse AM is in the *z*-direction. The *z* component of AM is a conserved quantity[4] $\mathbf{J}_z = \mathbf{L}_z + \mathbf{S}_z$ = constant, where $\mathbf{L}_z = \mathbf{R} \times \mathbf{P}$ is the orbital AM generated by the curved propagation of light[30]. Here $\mathbf{R} = \mathbf{r}/\kappa$ is the radius of the curved path and **P** is the momentum of light[31]. Since the momentum is altered by the real space geometric phase, the transverse spin AM must also be changed to satisfy AM conservation:

$$\begin{aligned}\mathbf{L}_z &= \mathbf{R} \times \mathbf{P} = \mathbf{R} \times (\mathbf{P}^0 + \delta\mathbf{P}) \\ \mathbf{S}_z &= \mathbf{S}_z^0 - \mathbf{R} \times \delta\mathbf{P}\end{aligned} \quad , \tag{3}$$

where $\mathbf{P}^0$ and $\mathbf{S}_z^0$ are the momentum and transverse spin AM in a flat path, respectively. And $\delta\mathbf{P} \propto \nabla_r \Phi_G \propto -\text{sgn}(\mathbf{s}_t \cdot \mathbf{l}_t)\kappa$ is the change of the momentum due to the geometric phase. The changing of transverse spin density depends on $\text{sgn}(\mathbf{s}_t \cdot \mathbf{l}_t)$ accordingly.

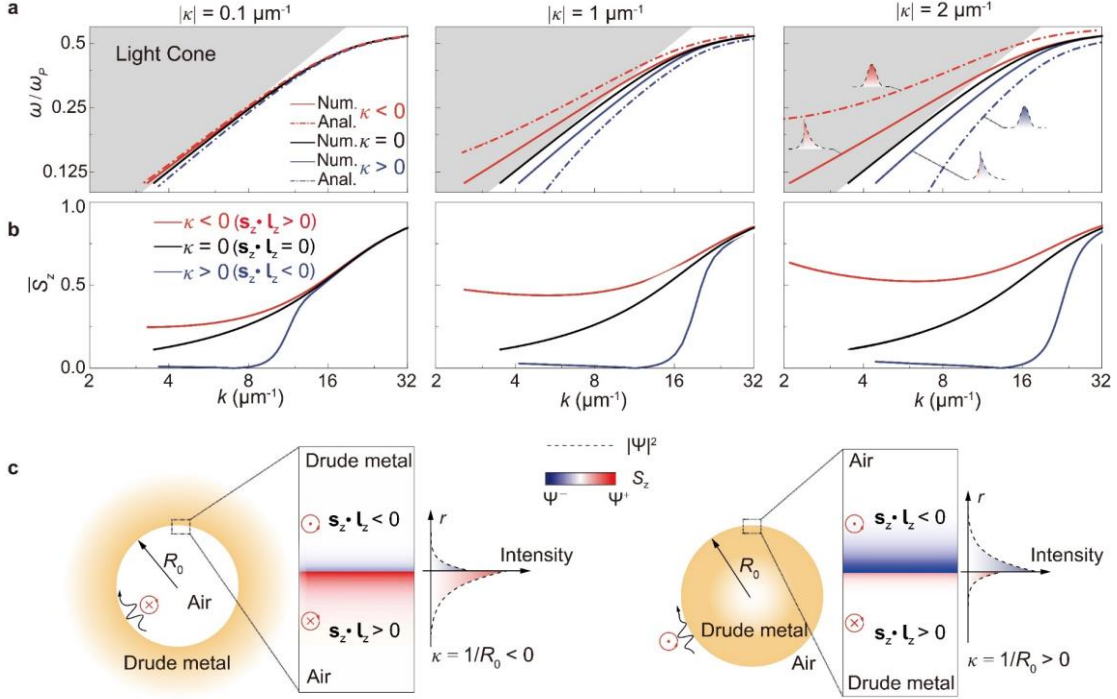

**Fig. 3. (a)** Dispersion relation and **(b)** Transverse spin of surface plasmon on the path with different curvature. **(c)** Distribution of the transverse spin density of plasmon. The simulation is carried out through eigenmode analysis. The material is chosen to be Drude metal with $\varepsilon(\omega) = 1 - \omega_p^2/\omega^2$. The integration of transverse spin was carried out in the air. The integration of transverse spin in metal is given in SM.

The solutions of Eq. (1) are in the form of $\mathbf{\psi}^\pm = C_1 e^{\pm i k_{\uparrow\downarrow} r} + C_2 e^{\mp i k_{\uparrow\uparrow} r}$, where $C_1$ and $C_2$ are the normalization coefficients. $k_{\uparrow\downarrow} = \sqrt{k^2 + \kappa^2/4} + \kappa$ and $k_{\uparrow\uparrow} = \sqrt{k^2 + \kappa^2/4} - \kappa$ are the wave vector modified by the transverse SOI term, where the subscript arrows denote the relative direction between spin and orbital AM. Written in a simpler form, the transverse SOI wavevector becomes $k_{\text{TSOI}} = \sqrt{k^2 + \kappa^2/4} - \text{sgn}(\mathbf{s}_t \cdot \mathbf{l}_t)\kappa$. As a result, the dispersion relation of light with transverse SOI is given by

$$E_{\text{TSOI}} = \hbar c \left( \sqrt{k^2 + \frac{\kappa^2}{4}} - \text{sgn}(\mathbf{s}_t \cdot \mathbf{l}_t)\kappa \right). \tag{4}$$

Here, $c$ is the speed of light. This equation indicates the splitting of the dispersion relation of the opposite transverse spin state due to transverse SOI as shown in Fig. 2c. The dispersion relation is strongly dependent on both the curvature of the path and the wave vector. The strong nonlinear effect is more obvious for larger curvature and lower wave number (Fig. 2c). When the curvature vanishes $\kappa = 0$, the dispersion relation is back to the linear dispersion of light $E = \hbar c k$.

To demonstrate the transverse SOI, we numerically consider a surface plasmon polariton propagating on a curved path (*xy*-plane) through COMSOL Multiphysics. The dispersion relation is obtained by mode analysis (Fig. S1). And the transverse spin density is integrated and normalized through $\overline{S}_t = \overline{S}_z = \omega \iint_{x,y} \mathbf{S} ds / \iint_{x,y} W ds$, where $W = \left( \tilde{\varepsilon} |\mathbf{E}|^2 + \tilde{\mu} |\mathbf{H}|^2 \right)/4$ is the energy density, and the spin density is defined as[31] $\mathbf{S} = \frac{1}{4\omega} \mathrm{Im}\left( \tilde{\varepsilon} \mathbf{E}^* \times \mathbf{E} + \tilde{\mu} \mathbf{H}^* \times \mathbf{H} \right)$, where $\tilde{\varepsilon} = \partial(\omega\varepsilon)/\partial\omega$ and $\tilde{\mu} = \partial(\omega\mu)/\partial\omega$ are modified permittivity and permeability of the medium. It is clear from Fig. 3a and 3b that there is a splitting of both the energy dispersion and the transverse spin for opposite curvatures of the path, fulfilling the conservation of AM (See Fig. S4). And the splitting of dispersion and transverse spin is more obvious for larger curvature and small wave vectors, as shown in Eq. 3. The dispersion relation located in the light cone are the bound state of the concave waveguide with $\kappa < 0$. In fact, such anomalous dispersion relation of surface plasmon polariton and large transverse spin density in a curved path has been observed previously[32], yet, had not been well explained. Such intrinsic transverse SOI is universal in evanescent waves including photonic or polaritonic systems. The tailoring of local spin density is of paramount importance for designing spin-controlled directionality waveguides[33]. Generally, the density of transverse spin is dependent only on the confinement of an evanescent wave[34]. Based on transverse SOI, geometrical curvature becomes a new degree of freedom to control the transverse spin[8,33]. Due to the coexistence of two sides for an interface, there are two transverse spin states in each evanescent wave[5,34]. The two transverse spins will have opposite transverse SOI in each evanescent wave, their effect cannot be ideally separated. As a result, the numerical results for surface plasmon polaritons are a derivation from the analytical ones (shown in Fig. 3c).

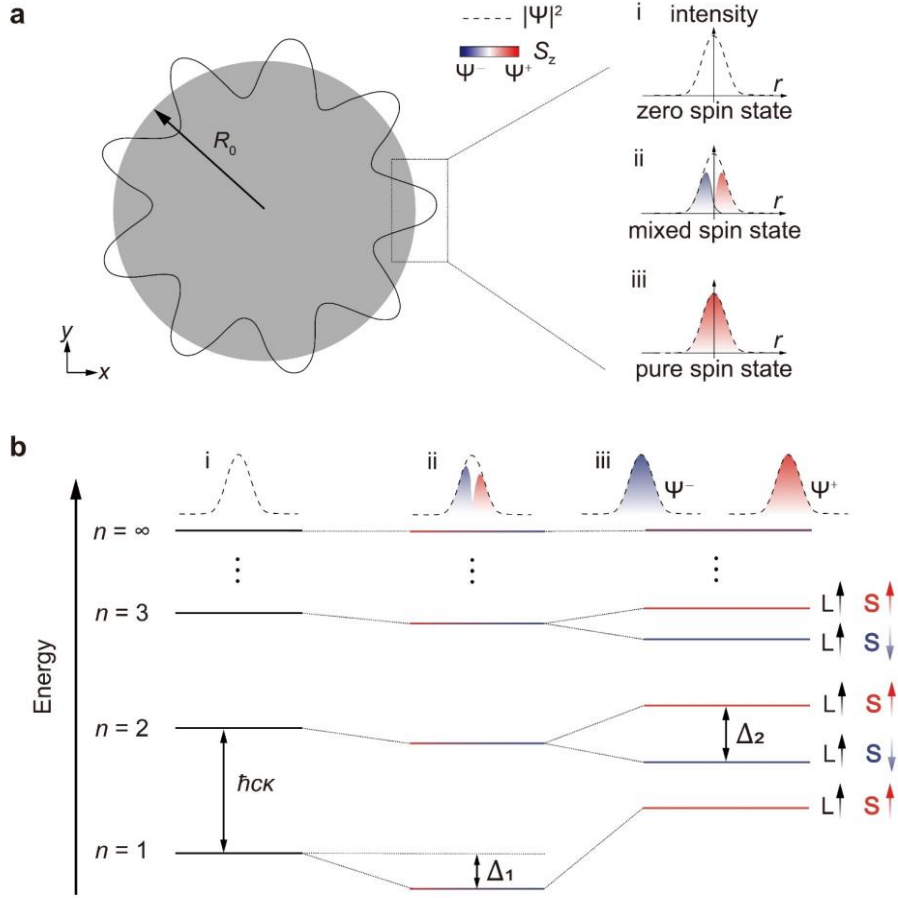

**Fig. 4. Fine structure of optical WGM. (a)** Schematic of optical WGM and transverse spin state. **(b)** The energy level of optical WGM with different transverse spin. Here, zero spin state (i) means WGM without considering the transverse SOI. The mixed spin state (ii) represent the state in real WGMs, which cannot be described by analytical results. The pure spin state (iii) denotes the state which can be described by analytical results. A pure spin state does not exist in a real system; it can only be realized by design. Here, the energy difference are $\Delta_1 \propto \hbar c \kappa \left( \alpha_1 \sqrt[3]{(n+1/2)} + \alpha_2 1/\sqrt[3]{(n+1/2)} - \alpha_3 1/^{2/3}\sqrt{(n+1/2)} \right)$, and $\Delta_2 = \hbar c \kappa \left( \sqrt{[n+1]^2 - 1/4} - \sqrt{[n-1]^2 - 1/4} \right)$, respectively.

The transverse SOI can also explain the energy levels of optical WGM, which is the resonance of the transverse spinning modes in circular cavities[35]. Considering an optical ring resonator with radius $R_0$ as shown in Fig. 4a. The energy levels of this ring resonator are $E_n = n\hbar c / R_0 = n\hbar c \kappa$, which can be obtained by solving the original

Helmholtz equation with a periodic condition or simply (Fig. 4b left panel): $kR_0 = n$, where $n \in \mathbb{Z}^+$ and $E = \hbar ck$ is the free space dispersion relation. Yet, as the coexistence of both transverse spin states discussed above, the analytical results cannot be obtained for real photonic WGM[36]. The effect of the transverse SOI, on the other hand, can be identified qualitatively from its asymptotic formulas: $k \propto \kappa(n + \alpha_1 \sqrt[3]{(n+1/2)} + \alpha_2 1/\sqrt[3]{(n+1/2)} - \alpha_3 1/^{2}\sqrt[3]{(n+1/2)})$, where $\alpha_i$ are parameters that are dependent on both the refractive index and the Airy function solution[36]. There are additional terms besides the ordinary term $\kappa n$. We argue that these terms are the corrections of momentum because of the transverse SOI. The second and third terms that have positive signs are contributed by transverse SOI with $\mathbf{s}_t \cdot \mathbf{l}_t < 0$, while the last term is contributed by transverse SOI with $\mathbf{s}_t \cdot \mathbf{l}_t > 0$. One can find that the correction of the effective wave vector is strong for lower energy (or low wave vector) ($n \to 1$), but weak for high energy ($n \to \infty$), which is consistent with our previous analysis. The correction of energy levels for real WGM is negative due to the energy is mainly distributed in the area with $\mathbf{s}_t \cdot \mathbf{l}_t < 0$ (middle panel of Fig. 4b).

As shown in the right panel of Fig. 4b, the energy level of WGM with a pure transverse spin can be obtained analytically by solving Eq. (1) with a periodic condition (Supplementary Section 4)

$$E_n^{\text{TSOI}} = \hbar c \kappa \sqrt{[n + \text{sgn}(\mathbf{s}_t \cdot \mathbf{l}_t)]^2 - 1/4}. \tag{5}$$

Notably, the energy structure of optical WGM with transverse SOI is the same as the fine structure of an atom. The energy level is higher for $\mathbf{s}_t \cdot \mathbf{l}_t > 0$ and lower for $\mathbf{s}_t \cdot \mathbf{l}_t < 0$ with respect to the original one $\mathbf{s}_t \cdot \mathbf{l}_t = 0$. As shown in Fig. 4b, when $\mathbf{s}_t \cdot \mathbf{l}_t < 0$, the energy level for $n = 1$ is missing ($\sqrt{[n + \text{sgn}(\mathbf{s}_t \cdot \mathbf{l}_t)]^2 - 1/4} = \sqrt{-1/4}$ is not exist at this particular parameter). The absence of the energy level for $n = 1$ reside in the scalar geometrical field ($\kappa^2/4$) in Eq. 1. Novel ideas are expected for testing the fine structure of optical WGM with the design of pure handedness of transverse spin[6,37].

# Conclusion

In conclusion, we have shown, for the first time, an intrinsic transverse SOI of light from the fundamental wave equation. Different from the longitudinal SOI, the transverse SOI of light can shift the dispersion relations in momentum space when a transversely spinning light propagates in a curved path and also the energy level of optical WGMs. These results on the one hand enable us to explain previous numerical and experimental results related to the evanescent wave in a curved path, on the other hand rise analogous between massless spin-1 particles and electrons. Furthermore, the intrinsic transverse SOI also provides us with another method to tailor the local density of transverse spin AM, which could be useful in chiral quantum optics and nanoscale chiral light-matter interactions.

The transverse SOI may appear in different types of guided waves, including but not limited to various polaritons, i.e., exciton-polariton, phonon polariton, etc. As a consequence, this geometric phase may provide a new paradigm to design integrated circuits and metasurfaces based on transverse SOI. According to the conservation of the AM, the transverse SOI can provide an effective way to tailor the local transverse spin density by the curvature of the waveguide. Since the gauge field of longitudinal and transverse SOI sustain in the momentum and the real space, respectively, they together provide a feasible way to realize the non-Abelian gauge field[38,39]. Given the universality of the conservation law of AM, these results could be extended to other classical waves that possess AM[40], either the spin or the orbital. For example, the surface waves in acoustic metamaterials with transverse spin[41] or the spatiotemporal optical vortices[42] travel in a curved path. The real space gauge field stemming from the curved space can also arise new schemes for astrophysics, such as stellar or gravitational wave detection.

Author contribution:

S.P.Z. and T.F. initiated the project. T.F. explains the transverse SOI. J.X.L. and T.F. carried out the numerical simulation. J.J.J. solved the analytical equation. T.F. and S.P.Z. wrote the manuscript. S.P.Z and H.X.X. supervised the project. All authors discussed and commented on the manuscript.

We thank Prof. Yuntian Chen, Prof. Li Mao, and Prof. Rui Yu for their helpful discussions and feedback. This work is supported by the National Natural Science Foundation of China (Grant No. 12134011), the National Key R&D Program of China (Grant No. 2021YFA1401104), and Key R&D Program of Hubei Province (2022BAA016), and the Fundamental Research Funds for the Central Universities. S. Z. is also supported by the Young Top-notch Talent for Ten Thousand Talent Program (2020-2023).

**Reference**

[1] G. Dresselhaus, Phys. Rev. **100**, 580 (1955).
[2] Y. A. Bychkov and É. I. Rashba, JETP Lett. **39**, 78 (1984).
[3] J. Sinova, S. O. Valenzuela, J. Wunderlich, C. Back, and T. Jungwirth, Rev. Mod. Phys. **87**, 1213 (2015).
[4] K. Y. Bliokh, F. J. Rodríguez-Fortuño, F. Nori, and A. V. Zayats, Nat. Photonics **9**, 796 (2015).
[5] K. Y. Bliokh and F. Nori, Phys. Rev. A **85**, 061801 (2012).
[6] L. Peng, L. Duan, K. Wang, F. Gao, L. Zhang, G. Wang, Y. Yang, H. Chen, and S. Zhang, Nat. Photonics **13**, 878 (2019).
[7] J. S. Eismann, L. H. Nicholls, D. J. Roth, M. A. Alonso, P. Banzer, F. J. Rodríguez-Fortuño, A. V. Zayats, F. Nori, and K. Y. Bliokh, Nat. Photonics **15**, 156 (2020).
[8] A. Aiello, P. Banzer, M. Neugebauer, and G. Leuchs, Nat. Photonics **9**, 789 (2015).
[9] K. Y. Bliokh and Y. P. Bliokh, Phys. Rev. Lett. **96**, 073903 (2006).
[10] O. Hosten and P. Kwiat, Science **319**, 787 (2008).
[11] K. Y. Bliokh, A. Niv, V. Kleiner, and E. Hasman, Nat. Photonics **2**, 748 (2008).
[12] Y. Zhao, J. S. Edgar, G. D. Jeffries, D. McGloin, and D. T. Chiu, Phys. Rev. Lett. **99**, 073901 (2007).
[13] N. Bokor, Y. Iketaki, T. Watanabe, and M. Fujii, Opt. Express **13**, 10440 (2005).
[14] C. Schwartz and A. Dogariu, Opt. Express **14**, 8425 (2006).
[15] T. A. Nieminen, A. B. Stilgoe, N. R. Heckenberg, and H. Rubinsztein-Dunlop, J. Opt. **10**, 115005 (2008).
[16] D. Haefner, S. Sukhov, and A. Dogariu, Phys. Rev. Lett. **102**, 123903 (2009).
[17] K. Y. Bliokh, E. A. Ostrovskaya, M. A. Alonso, O. G. Rodríguez-Herrera, D. Lara, and C. Dainty, Opt. Express **19**, 26132 (2011).
[18] L. Vuong, A. Adam, J. Brok, P. Planken, and H. Urbach, Phys. Rev. Lett. **104**, 083903 (2010).
[19] D. Pan, H. Wei, L. Gao, and H. Xu, Phys. Rev. Lett. **117**, 166803 (2016).
[20] N. Shitrit, I. Yulevich, E. Maguid, D. Ozeri, D. Veksler, V. Kleiner, and E. Hasman, Science **340**, 724 (2013).
[21] K. Rechcińska *et al.*, Science **366**, 727 (2019).
[22] X. Yin, Z. Ye, J. Rho, Y. Wang, and X. Zhang, Science **339**, 1405 (2013).
[23] L. Peng *et al.*, Sci. Adv. **8**, eabo6033 (2022).
[24] M.-Y. Lai, Y.-L. Wang, G.-H. Liang, and H.-S. Zong, Phys. Rev. A **100**, 033825 (2019).
[25] Z. Shao, J. Zhu, Y. Chen, Y. Zhang, and S. Yu, Nat. Commun. **9**, 926 (2018).
[26] L. Ma, S. Li, V. M. Fomin, M. Hentschel, J. B. Götte, Y. Yin, M. Jorgensen, and O. G. Schmidt, Nat. Commun. **7**, 10983 (2016).
[27] J. Wang *et al.*, Nat. Photonics **17**, 120 (2023).
[28] Y. Aharonov and D. Bohm, Phys. Rev. **115**, 485 (1959).
[29] M. V. Berry, Proc. R. Soc. Lond. A **392**, 45 (1984).
[30] L. Allen, M. W. Beijersbergen, R. J. C. Spreeuw, and J. P. Woerdman, Phys. Rev. A **45**, 8185 (1992).
[31] K. Y. Bliokh, A. Y. Bekshaev, and F. Nori, Phys. Rev. Lett. **119**, 073901 (2017).

[32] J.-W. Liaw and P.-T. Wu, Opt. Express **16**, 4945 (2008).
[33] B. Lang, D. P. S. McCutcheon, E. Harbord, A. B. Young, and R. Oulton, Phys. Rev. Lett. **128**, 073602 (2022).
[34] T. Van Mechelen and Z. Jacob, Optica **3**, 118 (2016).
[35] W. Chen, Ş. Kaya Özdemir, G. Zhao, J. Wiersig, and L. Yang, Nature **548**, 192 (2017).
[36] C. Lam, P. T. Leung, and K. Young, J. Opt. Soc. Am. B **9**, 1585 (1992).
[37] X. Piao, S. Yu, and N. Park, Phys. Rev. Lett. **120**, 203901 (2018).
[38] Y. Chen, R.-Y. Zhang, Z. Xiong, Z. H. Hang, J. Li, J. Q. Shen, and C. T. Chan, Nat. Commun. **10**, 1 (2019).
[39] Y. Yang, C. Peng, D. Zhu, H. Buljan, J. D. Joannopoulos, B. Zhen, and M. Soljačić, Science **365**, 1021 (2019).
[40] S. Wang, G. Zhang, X. Wang, Q. Tong, J. Li, and G. Ma, Nat. Commun. **12**, 6125 (2021).
[41] C. Shi, R. Zhao, Y. Long, S. Yang, Y. Wang, H. Chen, J. Ren, and X. Zhang, Natl. Sci. Rev., 707 (2019).
[42] A. Chong, C. Wan, J. Chen, and Q. Zhan, Nat. Photonics **14**, 350 (2020).